**Title:** Consequences of fleet diversification in managed and unmanaged fisheries

**Authors:** Matthew G. Burgess[a,b,c,d]

**Author Affiliations:**

[a]Department of Ecology, Evolution and Behavior, University of Minnesota, 1987 Upper Buford Circle, St Paul, Minnesota, 55108, USA.

Current Address: [b]Sustainable Fisheries Group, [c]Bren School of Environmental Science and Management, and [d]Marine Science Institute, University of California, Santa Barbara, CA 93106, USA

**Corresponding Author:**

Matthew G. Burgess

Sustainable Fisheries Group

Bren School of Environmental Science and Management & Marine Science Institute

University of California

Santa Barbara, CA 93106-5131

mburgess@ucsb.edu

Tel: 612-799-0208

**Abstract.**

Biological diversity is known to play an important role in generating and maintaining ecosystem productivity and other functions, and has consequently become a central focus of many efforts to preserve ecosystem services. Theoretical parallels suggest the diversity of fishing fleets may have a similarly important role in determining the productivity and ecological impacts of fisheries, but this possibility has rarely been explored. Here I present theoretical analyses showing that the diversity of métiers – combinations of technology, target species, and fishing grounds – and technical efficiencies in a fishing fleet have important impacts on the productivity, profitability, and ecological impacts of fisheries, particularly mixed-stock or multispecies fisheries. Diversification of métiers can increase yields and reduce threats to weak stocks in both managed and unmanaged multispecies fisheries. Diversification of technical efficiencies creates opportunities for larger profits in managed fisheries, but often decreases yields and worsens impacts on weak stocks in unmanaged fisheries. These results suggest that the potential impact of management may be highest in fisheries with diverse fleets.

**Introduction**

Biological diversity is widely considered to be an important driver of the productivity and stability of ecosystems and the services they provide (see McCann 2000; Loreau et al. 2001; Palumbi et al. 2009; Cardinale et al. 2012 for recent reviews). In ecological communities of consumers, diversity leads to high productivity through complementarity effects, whereby species with different resource requirements are able to partition resources more efficiently (Tilman et al. 1997; Lehman and Tilman 2000; Loreau and Hector 2001; Thébault and Loreau 2003); and sampling effects, whereby more diverse communities are more likely to contain highly productive species by chance (Tilman et al. 1997; Loreau and Hector 2001). Both of these effects cause consumer diversity to impact resource availabilities (Tilman et al. 1997). Diversity increases the stability of ecosystems' productivities and services through 'portfolio effects', whereby functional redundancies between species allow sudden perturbations affecting one species to be compensated by other similar species; and 'negative covariance effects', whereby sudden collapses in the abundance of a species lead to compensatory increases in the abundances of its competitors (Lehman and Tilman 2000; McCann 2000).

The role of biological diversity in promoting high and stable yields in fisheries has been well studied, with similar conclusions: Diverse aquatic ecosystems are generally thought to lead to more productive and stable fisheries (e.g. Worm et al. 2006; Palumbi et al. 2009; Schindler et al. 2010). The diversity of fishing fleets is also likely to play an important role in determining the productivity, stability, and ecological impacts of fisheries, but this has not been widely or systematically explored. Though there are strong theoretical parallels between fishers and ecological consumers, there are also some important differences. Fishers' behaviours and incentives can be complex and plastic, and can be influenced by management (Branch et al. 2006). Fisheries management can also have varying goals including profits, employment, food production, subsistence, recreation, and cultural value (Branch et al. 2006; Beddington et al. 2007; Worm et al. 2009; Chan et al. 2012). Fishers can vary widely in their technical efficiencies, and thus may not be bound to the same types of tradeoff surfaces as many other organisms are hypothesized to be – where it has been suggested that differences among species in overall resource-use efficiency are minor compared to differences in resource specialization (Tilman 2011).

Fleets can be diverse in a variety of ways. Fishers within a particular fishery may differ from one another in terms of their choice of fishing grounds, vessel sizes and gear-types, and target species; as well as their objectives, expert knowledge, behavioural plasticity, risk aversion, and other differences (Branch et al. 2005; 2006). Recent evidence from some fisheries suggests that certain types of fleet diversity can create important opportunities for progress toward management objectives in certain situations. For example, in mixed-stock or multispecies fisheries, in which any particular gear-type is likely to catch multiple species or stocks (populations), management often faces the challenge of navigating tradeoffs between overexploiting slow-growing populations, often termed 'weak stocks', and under-exploiting more rapidly reproducing populations (Hilborn 1976; Boyce 1996; Hilborn et al. 2004). Recent studies on multispecies trawl and groundfish fisheries in the E.U. (Marchal et al. 2011; Ulrich et al. 2011), United States (Dougherty et al. 2013), Canada and New Zealand (Sanchirico et al. 2006), and tuna fisheries in the Pacific (Sibert et al. 2012) have each found that management can overcome some of these tradeoffs at large spatial scales through management schemes that govern the allocation of fishing rights among fishers having different fishing grounds, gears or target species. In fisheries with diversity in economic efficiency, management using individual transferable quotas (ITQs) can increase fishery profits by encouraging the trading of fishing rights from low-efficiency fishers to high-efficiency fishers (Grafton et al. 2000; Marchal et al. 2011; Grainger and Costello 2012; Schnier and Felthoven 2013).

Here, I use a combination of general theoretical models and stochastic simulations to illustrate some important opportunities and pitfalls presented by fleet diversity to the yields, profits, and ecological objectives of fisheries. Following recent bioeconomic literature, I consider fleet diversity in terms of two concepts: 1) 'métiers' (ICES 2003; Marchal et al. 2013) or 'fishing opportunities' (Branch et al. 2005) – unique combinations of fishing gear, target species and geographic and temporal targeting that have roughly uniform relative catch-rates of different stocks in a multispecies or multi-stock fishery; and 2) technical efficiency (Hilborn 1985; Marchal et al. 2013) – which determines the costs at which fishers obtain particular catches at particular abundances.

I focus on two broad classes of fisheries: 1) those in which collective action towards fishery-wide goals is possible; and 2) fisheries in which there is neither management nor access restrictions, and fishery-wide outcomes are thus shaped predominantly by uncooperative competition among fishers (Clark 1976). For brevity, I hereafter refer to the former type of fishery as 'managed' and the latter type of fishery as 'unmanaged', but note that not all unmanaged fisheries are open-access and uncooperative, nor is management necessarily a requisite of fishery-wide cooperation. In managed fisheries, I focus on the effects of fleet diversity on the range of achievable outcomes relative to common fishery-wide goals (yield- and profit-maximization, and weak stock protection), rather than comparing specific types of management (e.g. quotas, reserves, maritime zoning, etc.). In unmanaged fisheries, I focus on effects of fleet diversity on fishery-wide outcomes (yield, ecological impacts) via competition among fishers.

I provide theoretical evidence suggesting that: i) diversifying métiers in multispecies fisheries should often lead to higher yields and less overfishing of weak stocks in both managed and unmanaged fisheries – by creating opportunities for management to reduce tradeoffs between objectives for different stocks, and promoting balanced exploitation in unmanaged fisheries; and ii) diversifying technical efficiencies should create greater opportunities for management to increase fishery-wide profits, but lead to more severe depletions (and thus lower yields) in unmanaged fisheries. For brevity, my analysis focuses on equilibrium statics, and thus effects of fleet diversity on the temporal stability of yields, profits, stocks, and ecosystems, as well as other possible dynamic effects are not considered explicitly, but are discussed and deserve future study.

**Mathematical framework and operational definition of fleet diversity**

The analyses presented here use a 'fishing unit', defined as a group of fishers from a particular métier having a particular efficiency, as the base unit of fleet diversity. Métiers are defined in this analysis by their relative catch-rates of different stocks. Efficiency is defined by the cost of obtaining catch, controlling for relative catch rates and stock abundances. Thus, a particular fishing unit's métier is defined by relative catchabilities (per-capita-per-unit-effort catch rates) for different stocks, while its efficiency is determined in combination by the magnitudes of stocks' catchabilities and the per-unit-effort costs of fishing.

Defining fishing effort in terms of unit costs (i.e. measuring effort in dollars instead of hooks, days, etc.) allows for a simple mathematical representation of métiers and efficiency as components of catchability. Suppose $S$ different fish stocks are caught in a fishery, and each stock has population size, $N_i(t)$ for stock $i$ at time $t$, and a per-capita growth rate, $g_i(\mathbf{N}(t))$ for stock $i$, in the absence of fishing, where $\mathbf{N}(t) = (N_1(t), N_2(t), ..., N_S(t))$ (i.e. stocks may interact ecologically with one another). Suppose fishers in a particular fishing unit, $j$, are spending $E_j(t)$ (in dollars or some other monetary unit) on fishing effort and catch a fraction, $q_{ij}$, of individuals of stock $i$ with each dollar spent ($q_{ij}$ is the catchability of stock $i$ in fishing unit $j$). Because relative catch-rates are the key distinguishing features of fishing units for the purposes of this analysis, it is instructive to consider $q_{ij}$ constant for all stocks ($i$) and fishing units ($j$), and to consider changes in $q_{ij}$ for any particular fisher – resulting from catch hyperstability (Harley et al. 2001), economies of scale (e.g. Squires and Kirkley 1991), or changing behaviour, for example – as transitions between fishing units. Thus, stock $i$, if only targeted by fishing unit, $j$, has a rate of change, $dN_i(t)/dt$ at time $t$, given by:

$$\frac{dN_i(t)}{dt} = N_i(t)\left(g_i(\mathbf{N}(t)) - q_{ij}E_j(t)\right) \quad (1a).$$

The price-per-individual caught of stock $i$ is denoted $p_i$, and I allow it to be a function of abundance ($N_i$) (i.e. $p_i = p_i(N_i(t))$). I assume throughout this analysis that $p_i$ is either constant or increasing as the harvest rate of stock $i$ falls (i.e. $p_i'(.) \leq 0$) because harvest rate falls as population size (*N*) falls, and decreased harvest rate implies decreased supply, which typically leads to either constant or increasing prices depending on the availability of substitutes and elasticity of demand (Clark 1976, 1985). However, I also assume that price does not increase fast enough to result in rising per-unit-effort revenues with decreasing abundance (i.e. $d[p_i(N_i)N_i]/dN_i > 0$ for all $N_i$, $i$ is assumed). Violation of this assumption would likely result in the extinction of stock $i$ because profit margins would increase as abundance declined (see Appendix A; and Courchamp et al. 2006). Lastly, I assume that all fishers face the same per-individual price for each stock (i.e. $p_i(N_i(t))$ is the same for all fishing units) for simplicity, but some possible implications of violations of this assumption are discussed. The total profits of fishers in fishing unit $j$ at time $t$, $\pi_j(t)$, are thus given by,

$$\pi_j(t) = E_j(t)\left(\sum_{i=1}^{S} p_i(N_i(t))q_{ij}N_i(t) - 1\right) \quad (1b).$$

Métiers and efficiency are defined mathematically by partitioning $q_{ij}$ into two components: $m_{ij}$ (where $\sum_i m_{ij} = 1$), a fraction representing the métier's catch rate of stock $i$ relative to other stocks, and $e_j$, another constant representing fishing unit $j$'s efficiency, such that $q_{ij} = m_{ij}e_j$. With this substitution, equations (1a) and (1b) become:

$$\frac{dN_i(t)}{dt} = N_i(t)\left(g_i(\mathbf{N}(t)) - m_{ij}e_j E_j(t)\right) \quad (2a)$$

$$\pi_j(t) = E_j(t)\left(e_j \sum_{i=1}^{S} p_i(N_i(t))m_{ij}N_i(t) - 1\right) \quad (2b).$$

It can be seen from equations (2a) and (2b) that $\{m_{ij}\}_{\forall i}$ captures the relative catch rates of different stocks by fishers in fishing unit $j$ (i.e. its métier), and $e_j$ captures the cost efficiency of catch in fishing unit $j$, controlling for relative catch rates and stock abundances (i.e. its technical efficiency). With multiple fishing units in the fishery (where the number of fishing units is denoted *J*), the rate of change of the abundance of stock *i* is given by:

$$\frac{dN_i(t)}{dt} = N_i(t)\left(g_i(\mathbf{N}(t)) - \sum_{j=1}^{J} m_{ij} e_j E_j(t)\right) \quad (2c).$$

At equilibrium, $dN_i(t)/dt = 0$ for all *i*, and $dE_j(t)/dt = 0$ for all *j*. For stock *i*, this means that:

$$N_i^* = 0 \text{ or } g_i(\mathbf{N}^*) = \sum_{j=1}^{J} m_{ij} e_j E_j^* \quad (3),$$

(by equation 2c), where $N_i^*$ is the equilibrium abundance of stock *i*, $\mathbf{N}^* = \{N_1^*,\ldots,N_S^*\}$ and $E_j^*$ is the equilibrium fishing effort in fishing unit *j*. Assumptions about the determinants of effort levels depend on the management context. In managed fisheries, I assume effort levels (i.e. $E_j(t)$ and $E_j^*$ for all *j*) are controllable by managers. The range of possible equilibria is determined by the diversity in the métiers ($\{\{m_{ij}\}_{\forall i}\}_{\forall j}$) and efficiencies ($\{e_j\}_{\forall j}$) of available fishing units in the fleet. Equation (3) represents the manner in which this range is constrained by fleet diversity and by the ecology of the stocks (represented by $g_i(\mathbf{N}^*)$ for each stock *i*). The specific equilibrium in any managed fishery will be determined by this available range of possible equilibria, the management objective (e.g. yields, profits, etc.), and the degree to which efforts can be controlled by the managers. My analysis of managed fisheries focuses on exploring the effects of fleet diversity on the ranges of different management objectives that are technologically and ecologically achievable. To do this, I assume that $E_j$'s are completely controllable, and thus do not compare specific management instruments. In unmanaged fisheries, I assume that effort levels respond to profits, and equilibrium outcomes are ultimately determined by competition, as discussed in detail in the relevant section below.

**Fleet diversity in managed fisheries**

In managed fisheries, fleet diversity provides management with more degrees of freedom, which increases the achievable fishery-wide yields and profits, and creates opportunities to reduce tradeoffs between fishery yields and profits and minimizing adverse impacts on weak stocks and other species. Specifically, diversity in efficiency creates opportunities for management to increase fishery-wide profits by implementing a policy that causes the most efficient fishers to take larger shares of the catch. Higher diversity in efficiency also increases the likelihood of having high-efficiency extremes in the fishery, as a result of a sampling effect (see Appendix B for mathematical proof; Tilman et al. 1997 for analogous proof of the sampling effect of biodiversity on ecosystem productivity). Diversity in métiers in multispecies fisheries expands the spectrum of possible relative aggregate catch rates of different stocks, which increases achievable fishery-wide yields and profits, and creates opportunities for management to reconcile profit-and yield-maximization with prevention of overfishing and other ecological objectives (Figure 1, Figure B1; see Appendix B for expanded mathematical treatment).

With few different métiers in a multispecies fishery, profits, yields, and management goals for different stocks often tradeoff with one another as a result of fishers' lack of control over the relative catch-rates of different stocks in the fishery (Boyce 1996; Squires et al. 1998) (Figure 1a,b). As a result, it is often impossible to set a target catch or effort-quota that does not either over-exploit some stocks or under-exploit others (Figure 1b). However, a multispecies fishery with multiple métiers that are diverse in their relative catch rates can achieve a much wider range of combinations of exploitation rates and equilibrium abundances of its stocks. This is accomplished by influencing both total fishing effort and relative fishing efforts among métiers through management (Figure 1c; see Appendix B).

To be precise, in a multispecies fishery with only a single métier, $j$, the set of possible equilibrium abundances of different stocks is constrained to a one-dimensional hypersurface (i.e. a curve), defined by equation (3), whose shape is largely determined by the stocks' relative catch rates in the métier and their relative population growth rates (Figure 1a,b; Burgess et al. 2013). I hereafter refer to this as the 'vulnerability constraint' of métier $j$, because the relative catch rates and population growth rates of stocks determine their relative 'vulnerabilities' to depletion by métier $j$ (Burgess et al. 2013). For example, if all stocks have logistic population growth ($g_i(.) = r_i(1 - (N_i/K_i))$ for all $i$, where $r_i$ is the maximum per-capita growth rate and $K_i$ is the carrying capacity of stock $i$ (Schaefer 1954)), then the equilibrium abundances, $N_x^*$ and $N_y^*$, of any two non-extinct stocks, $x$ and $y$, must satisfy:

$$\frac{\left(1 - \frac{N_x^*}{K_x}\right)}{\left(1 - \frac{N_y^*}{K_y}\right)} = \frac{\left(m_{xj}/r_x\right)}{\left(m_{yj}/r_y\right)} \quad (4).$$

Equation (4) is derived by combining equation (3) for both stock $x$ and stock $y$, under the assumptions of logistic growth and a single fishing unit, $j$; and defines the vulnerability constraint of métier $j$ in this logistic model, which is linear with a slope determined by the relative values of $m_{ij}/r_i$ (normalized catchability/maximum growth rate) for different stocks (see also Holt 1977; Clark 1985). The shapes of vulnerability constraints under some other types of ecological assumptions are discussed in Appendix A, and illustrated in Figure A1a-e.

In a multispecies fishery with multiple métiers, any desired combination of stocks' equilibrium abundances that lies in the region in population space bounded by the vulnerability constraints of the different métiers could be achieved by implementing a policy or management strategy that influences both the total effort fishery-wide and relative efforts in different métiers (Figure 1c). This is illustrated mathematically in Appendix B. The space of possible sets of equilibrium stock abundances in a fishery has a dimensionality determined by the number of different métiers. Thus, it is highly unlikely that a particular target set of stock abundances in a managed fishery will be achievable if there are fewer métiers than stocks to be managed. This is also shown mathematically in Appendix B, and illustrated graphically in Figure B1. Additionally, because the relative impacts of a métier on different stocks are determined by the relative vulnerabilities of stocks to the métier ($m_{ij}/r_i$ in the logistic model) rather than simply their relative catch rates ($m_{ij}$), it is the diversity in relative vulnerabilities of stocks among métiers (i.e. differences in $\{(m_{ij}/r_i)\}_{vi}$ rather than $\{m_{ij}\}_{vi}$) that is particularly important in providing opportunities for yield and profit gains in managed fisheries.

Figure 2 shows the results of stochastic simulations of a 5-stock fishery illustrating the opportunities that fleet diversity offers managed fisheries for increases in yields and profits, and reduction in weak stock collapses by avoiding the inter-stock tradeoffs common in multispecies fisheries. For simplicity, each stock is assumed to have logistic population growth and constant prices, though the qualitative results generalize to more complex models. Each simulation fixes the number of fishing units and randomly generates 500 parameter sets, $\{r_i, m_{ij}, e_j, K_i\}$ for all $i$ and $j$ ($r_i$, ($m_{ij}/r_i$) ~ U[0,1] ($m_{ij}$'s are normalized after each draw to sum to 1); $e_j$ ~ U[1,10]; $K_i$ ~ U[10,100]; for simplicity, $p_i$ is fixed at 1 for all stocks, as $K_i$ already provides a randomly selected determinant of the relative values of the same per-capita catch rate from different stocks), and sets equilibrium fishing efforts, $E_j^*$ for all $j$, in order to maximize either yield (Figure 2a) or profit (Figure 2b) from all 5 stocks combined. The maximum achievable yield (MAY) or profit (MAP) with each parameter set is compared to the theoretical maximum yield (MTY; the sum of maximum sustainable yields (MSY) for all stocks) or profit (MTP) achievable given the stocks' ecological parameters and the bounds on efficiency placed on the random selection of fishing units. The average number of stocks persisting at the achievable maximum (MAY or MAP) is also reported. This procedure was repeated for both yield- (Figure 2a) and profit-maximization (Figure 2b), allowing métiers only, efficiencies only, or both to vary among fishing units within each random draw.

As the theory predicts, diversifying métiers increased the average maximum achievable yields (Figure 2a) and profits (Figure 2b). Diversifying métiers also reduced the average frequency of stock extinctions required for yield- or profit-maximization (Figure 2) by creating opportunities for management to avoid tradeoffs between overexploiting some stocks and under-exploiting others by influencing relative effort allocations among métiers. Diversifying efficiency increased achievable profits (Figure 2b), but had no effect on the achievability of yields (Figure 2a) or stock extinction frequencies at the optima (Figure 2). Results are qualitatively similar with $m_{ij}$ and $r_i$ drawn independently, but yields and profits increase more slowly with diversity of métiers, and saturate at higher diversities.

**Fleet diversity in unmanaged fisheries**

In an unmanaged fishery, effort levels are driven by profits and eventually determined by the conditions that make further effort unprofitable (Clark 1976) (Figure 3a). The impacts of fleet diversity on yields, profits, and stocks (Figure 4) are mediated by competition between fishers (Figure 3b,c,d). Competition tends to favour the most efficient fishers in the fishery (Figure 3b), and the likelihood of having high efficiency extremes in a fishing fleet increases with its diversity via a sampling effect (see Appendix B; Tilman et al. 1997). As a result, fleet diversification tends to lead to greater aggregate efficiency (Figure 4a), which allows profits at lower stock abundances and thereby often leads to decreases in long-term yields and increases in the frequency and severity of stock collapses (Figure 4b). Competition among fishers in different métiers tends to either favour métiers with more balanced exploitation rates (Figure 3c) or result in co-existence (Figure 3d), both of which lead to more balanced aggregate exploitation rates (Figure 4a), often resulting in higher yields and fewer stock collapses (Figure 4b). Fleet diversity has little to no effect on long-term profits, as they tend towards zero at any diversity of fishing units, provided there is no monopoly or oligopoly in the fleet's ownership (Clark 1976).

To illustrate these points graphically (Figures 3 and 4) and mathematically (see also Appendix C), I assume (following Clark (1976; 1985)) that fishing effort within an individual fishing unit increases when profits are positive, decreases when they are negative, and stays constant when profits are 0 (i.e. revenues are exactly equal to opportunity costs). In other words, I assume, for fishing unit $j$, that $dE_j(t)/dt < 0$ if $\pi_j(t) < 0$, $dE_j(t)/dt > 0$ if $\pi_j(t) > 0$, and $dE_j(t)/dt = 0$ if $\pi_j(t) = 0$. I do not make specific assumptions about the rate or manner in which effort adjusts to profit conditions. Thus, in a fishery catching $S$ stocks with only a single fishing unit, $j$, equilibrium would occur at a set of stock abundances, $\mathbf{N}^{*j} = \{N_1^{*j},\ldots,N_S^{*j}\}$, satisfying:

$$e_j \sum_{i=1}^{S} p_i(N_i^{*j}) m_{ij} N_i^{*j} = 1 \quad (5).$$

Equation (5) (derived by setting $\pi_j(t) = 0$ in equation (2b)) defines an $S - 1$ dimensional surface on which $\mathbf{N}^{*j}$ must lie, which I hereafter refer to as the 'profitability constraint' of fishing unit $j$. Similarly to the vulnerability constraint, the term 'profitability constraint' here refers to the fact that equilibrium abundances are constrained to this surface, rather than referring to other common uses of the term 'constraint' in mathematics (e.g. in optimization or control theory). The profitability constraint is illustrated in Figure 3 under the assumption of constant prices (where it is linear) and in Figure A1f under the assumption of increasing prices with decreasing abundance (where it is generally convex, see Appendix A). Its slope is determined by the métier (Figure 3c,d) and its position relative to the origin is determined by efficiency (Figure 3b). Equilibrium with a single fishing unit occurs at the intersection of its vulnerability and profitability constraints (Figure 3a) – where all $dN_i(t)/dt = 0$, and $dE_j(t)/dt = 0$ because $\pi_j(t) = 0$.

Competition between fishing units for fish has strong parallels with ecological communities of consumers competing for resources, which have been extensively studied. One of the seminal results in ecological competition theory is that outcomes of competition depend largely on species' abilities to invade communities of their

competitors – meaning they have positive growth rates when they are rare and competitors are established (MacArthur and Levins 1964; Levin 1970; Tilman 1980). As Tilman (1980) illustrates graphically and mathematically, if two species are competing and: i) each can invade the other's equilibrium, they co-exist; ii) one can invade the other's equilibrium, but not vice versa, the successful invader will competitively exclude the other; iii) neither can invade the other's equilibrium, one species will exclude the other, but which wins will depend on which establishes first or increases in population faster (called a 'priority effect'). The same principles apply to competing fishing units, except that it turns out that priority effects require fishers from pairs of fishing units to receive oppositely differing prices for the same catch of at least one pair of stocks (e.g. fishing unit $j$ receives a higher price than fishing unit $k$ for stock $x$ but a lower price for stock $y$; see Appendix C, Figure C1). Thus, priority effects may be relatively uncommon. There are some additional complexities when equilibria are not stable (e.g. see McGehee and Armstrong 1977; Armstrong and McGehee 1980), which I do not consider explicitly here for brevity, but do not affect the general principle that outcomes of competition are driven by species' (or analogously, fishing units') abilities to invade each other's established populations (Armstrong and McGehee 1980), and thus should also not affect the qualitative results I present concerning the effects of diversity of métiers and efficiency on yields and ecological outcomes.

As illustrated in Figure 3b, competition favours efficiency because more efficient fishers can still make profits at stock abundances resulting in zero profits for less efficient fishers within the same métier. This can be easily shown by substituting $e_k$ ($e_k > e_j$) into equation (5) for $e_j$, which would transform the equation to an inequality (i.e. $\pi_k(\mathbf{N}^{*j}) > 0$).

Similarly, less efficient fishers make negative profits at the equilibria of more efficient fishers of the same métier ($\pi_j(\mathbf{N}^{*k}) < 0$), which would eventually force them to exit the fishery. Because competition favours efficiency, and because increasing diversity increases the likelihood of including high efficiency extremes (see Appendix B), increasing fleet diversity should increase the aggregate efficiency of unmanaged fleets, on average (see Appendix C).

A high diversity of métiers is more likely to lead to balanced exploitation of the stocks in a fishery than a low diversity of métiers for two reasons: 1) competition among fishers with equal efficiency tends to favour those in métiers with more balanced exploitation of the different stocks (Figure 3c); and 2) competitive co-existence between fishers in two different métiers leads to aggregate relative exploitation rates among stocks that are intermediate to those that would be produced by each métier individually (Figure 3d). Specifically, in a fishery in which all fishing units have the same efficiency and all $m_{ij}$ between 0 and 1 are technologically feasible for any stock, $i$, (i.e. all possible métiers are feasible) competition among an infinitely diverse initial pool of métiers (i.e. a pool including all possible métiers) results in relative equilibrium abundances, $N_x^*$ and $N_y^*$, of any two extant stocks, $x$ and $y$, satisfying (see Appendix C):

$$p_x(N_x^*)N_x^* = p_y(N_y^*)N_y^* \quad (6).$$

In other words, all extant stocks have relative equilibrium abundances equal to the inverse of their equilibrium price-ratios (i.e. $(p_x^*/p_y^*) = (N_y^*/N_x^*)$, where $p_i^* = p_i(N_i^*)$).

This occurs because: i) in a fishery in which all fishers are equally efficient, there exists a métier that can invade and disrupt any equilibrium point not satisfying equation (6) (e.g. métier 2 in Figure 3c); and ii) co-existence among multiple fishing units occurs at equilibrium stock sizes at the intersection their profitability constraints (Figure 3d), and all possible profitability constraints with a particular efficiency intersect at a single point, at which equation (6) is satisfied (Figure 3c,d). Point ii) can be easily derived from equations (5) and (6) (see Appendix C). To illustrate point i), suppose a fishery in which all fishers have the same efficiency is at an equilibrium, $\mathbf{N}^* = \{N_1^*,…, N_S^*\}$, not satisfying equation (6). This implies that there is at least one pair of stocks, $x$ and $y$, with the property $p_x(N_x^*)N_x^* > p_y(N_y^*)N_y^*$. Given this fact, any new fishing unit, $k$, with the property relative to any established fishing unit, $j$, that $m_{xk} > m_{xj}$, $m_{yk} < m_{yj}$, and $m_{ik} = m_{ij}$ for all $i \neq x,y$, will have positive profits at this equilibrium, and thus be able to invade the fishery. Moreover, the invasion of fishing unit $k$ would increase the overall mortality rate of stock $x$ relative to stock $y$, which would reduce $N_x^*$ relative to $N_y^*$. Assuming that $d[p_i(N_i^*)N_i^*]/dN_i^* > 0$ ($i = x, y$) (i.e. revenues from any stock are positively related to its abundance), this would decrease the difference between $p_x(N_x^*)N_x^*$ and $p_y(N_y^*)N_y^*$, which would iteratively lead to equation (6) being satisfied if diversity increased infinitely (i.e. if all possible métiers were given the opportunity to try to invade the fishery).

Putting the results concerning efficiency and balanced exploitation together: In an unmanaged fishery, where all métiers are feasible for any efficiency up to a maximum efficiency, $e_{MAX}$, and métiers are independent of efficiencies, infinite initial fleet diversity results in an equilibrium, denoted $\mathbf{N}^{**e_{MAX}}$, at which equation (6) holds for all extant stocks, and all persisting fishing units have efficiency $e_{MAX}$ (see Appendix C for expanded proof). To be precise, $\mathbf{N}^{**e_{MAX}} = \{N_1^{**e_{MAX}},...,N_S^{**e_{MAX}}\}$ solves (from equations (5) and (6)),

$$p_i\left(N_i^{**e_{MAX}}\right)N_i^{**e_{MAX}} = 1/e_{MAX} \quad (7),$$

for all extant stocks. This is illustrated in a stochastic simulation of a fishery targeting two stocks having logistic growth and constant prices in Figure 4a (see caption for parameter values/distributions). As fleet diversity increases, the distribution of equilibria concentrates at $\mathbf{N}^{**e_{MAX}}$ (Figure 4a).

Of course, the convergence of stock sizes, as a result of fleet diversification, to the equilibrium, $\mathbf{N}^{**e_{MAX}}$, described by equation (7), depends on the assumption that efficiencies are distributed independently from métiers (i.e. $e_{MAX}$ is the same for all métiers). In reality, however, this is not likely to be the case. For example, if some stocks are generally easier or cheaper to catch than others due to their range or ecology, métiers with higher relative catch rates of these stocks are likely to be more efficient. One simple way to consider this in the modeling framework presented here is to let $q_{ij} = e_j a_i m_{ij}$, where $e_j$ measures the overall efficiency of fishing unit $j$ that is independent of its target stock, and $a_i$ ($a_i > 0$) measures how easy stock $i$ is to catch relative to other stocks. With this definition of $q_{ij}$, increasing fleet diversity in a fishery with maximum efficiency, $e_{MAX}$, will drive equilibrium stock abundances towards an equilibrium, $\mathbf{N}^{**e_{MAX}}$, described by:

$$p_i\left(N_i^{**e_{MAX}}\right) a_i N_i^{**e_{MAX}} = 1/e_{MAX} \quad (8),$$

where $p_x\left(N_x^{**e_{MAX}}\right) a_x N_x^{**e_{MAX}} = p_y\left(N_y^{**e_{MAX}}\right) a_y N_y^{**e_{MAX}}$ for any two stocks, $x$ and $y$. Equation (8) is derived using identical logic as equation (7) (see Appendix C). The general qualitative result is that diversification of métiers will tend to drive stocks towards relative abundances at which they generate equal revenue per-dollar spent on effort for equally efficient classes of technology, and diversification in efficiency will tend to broadly decrease stocks' abundances.

By promoting balanced exploitation, diversification of métiers will often lead to higher yields and reduce the likelihood of weak stock collapses. In contrast, diversification of efficiency will often reduce yields and increase the likelihood of weak stock collapses in by promoting high aggregate efficiency. Figure 4b illustrates an example of such patterns in a stochastic simulation of the same 2-stock fishery as in Figure 4a (see caption for parameter values/distributions).

Provided efficiency is finite, $\mathbf{N}^{**e_{\mathrm{MAX}}}$ will occur at a positive abundance for each commercially valued stock (i.e. each stock with a positive price) (by equation (7)/(8), see Figure 4a). Thus, the effect diversifying métiers reducing the likelihood of weak stock collapses will often dominate the opposite effect of diversifying efficiency at high diversities of both (because $N_i^{**e_{\mathrm{MAX}}} > 0$ for all $i$), meaning that broad increases in fleet diversity should often reduce the threat of weak stock collapses, as is the case in the simulated example in Figure 4b. However, whether the positive effect of diversifying métiers on equilibrium yield dominates the negative effect of diversifying efficiency will be much more context-dependent – driven largely by the maximum feasible efficiency, $e_{\mathrm{MAX}}$. If $e_{\mathrm{MAX}}$ is sufficiently large, as in the simulated example in Figure 4b, the effect of métier diversification increasing yield may dominate at very low fleet diversity, but the efficiency effect may dominate at higher diversity as the métier effect saturates. However, it is possible for the effect of diversifying métiers to dominate at high fleet diversity if $e_{\mathrm{MAX}}$ is sufficiently small.

While these results likely generalize to many types of fleets and fished stocks, there are some important exceptions. In particular, as discussed in Appendix D and illustrated in Figure D1, diversification in either métiers or efficiency often leads to the collapse of stocks whose non-substitutable prey or mutualists are also caught in the fishery; and diversifying métiers can sometimes increase the likelihood of stock collapses in fisheries where technological limitations make some relative catch-rates infeasible. Additionally, any stock whose price can rise fast enough to increase the revenues it generates as its abundance falls (i.e. $d[p_i(N_i)N_i]/dN_i < 0$ for stock $i$) is likely to be fished to extinction in general (Courchamp et al. 2006), but diversifying métiers or efficiency can also increase the chances of this (see Appendix A).

For by-catch populations, having little or no commercial value, the effect of diversifying métiers on the likelihood of collapse depends on the range of feasible relative catch rates and the way in which by-catch rates impact the efficiency of catching commercially valued stocks. If by-catch comes at an efficiency cost, then diversifying métiers could reduce impacts on by-catch species, as low-by-catch métiers would be favoured by competition. In contrast, if by-catch mitigation comes at an efficiency cost, then diversifying métiers could have the opposite effect, increasing the impacts on by-catch species. Diversifying métiers could similarly increase impacts on by-catch species if low-by-catch technologies were infeasible.

**Discussion**

This study presents two broad theoretical results: I) Diversifying métiers in multispecies fisheries often leads to higher yields and less overfishing in both managed and unmanaged fisheries – by creating opportunities to reduce tradeoffs between management objectives for different stocks and encouraging balanced exploitation in unmanaged fisheries. II) Diversifying technical efficiency creates opportunities for greater profits in managed fisheries, but tends to lead to more severe depletions (and thus lower yields) in unmanaged fisheries. Together, these results suggest that the potential value of transitioning to management in a fishery often increases with the diversity of its fleet.

The results concerning managed fisheries follow directly from the principle that more diverse fleets give managers more possible combinations of relative catch rates and costs (i.e. more degrees of freedom), with which they can more effectively balance objectives for different stocks and minimize costs, through controls influencing both total fishery-wide effort and relative efforts among different types of fishers. The results concerning unmanaged fisheries follow from the following principles of competition among fishers: i) All else equal, fishers with high technical efficiency (i.e. they obtain catches low costs relative to other fishers, with stock abundances being equal) will tend to outcompete less efficient fishers, and high efficiency extremes are more likely to be found in a diverse fishery. ii) If a particular fishing fleet is exploiting commercially valued stocks highly asymmetrically relative to their prices and abundances, and technology exists to adopt an alternate fishing practice that better targets a currently underexploited stock, someone will eventually adopt this practice and profitably enter the fishery, increasing the overall fishing pressure on the previously underexploited stock relative to others (i.e. balancing exploitation) as a result. Greater diversity in fishing technology increases the likelihood that technology will exist to exploit such economic opportunities. iii) Métiers with highly different catch profiles (relative catch-rates among stocks) compete less, and are likely to coexist in a fishery, resulting in relative aggregate catch rates among stocks that are intermediate to those each métier would produce on its own (e.g. Figure 3d). Greater diversity in métiers increases the likelihood of finding catch profiles across the full range in the fishery, thus more likely resulting in balanced aggregate exploitation.

Though these principles, and the results (I) and II)) they imply, are illustrated mathematically in this paper using simplifying assumptions that may be unrealistic – notably: a) that fleet diversity can be partitioned into discrete fishing units, each with uniform relative catch rates and efficiency; b) that unmanaged fishery equilibria are stable enough for basic equilibrium-based competition results in theoretical ecology (Tilman 1980; see Armstrong and McGehee 1980) to hold; and c) all fishers face the same prices for catch of each stock – they are likely to be more general. Similar theoretical simplifications have been used, for example, to conceptually illustrate some of the mechanisms underlying positive effects of biodiversity on ecosystem productivity (e.g. Tilman et al. 1997), which have proven to be robust both theoretically and empirically to many added complexities (see Cardinale et al. 2006, 2011 respectively for recent meta-analysis and review).

Thus, these model simplifications (a), b) and c) above) should be thought of as conceptual tools for understanding broad results, rather than accurate descriptions of reality. It is likely that some unmanaged fisheries have unstable or cyclic dynamics. Relative catch rates – the defining characteristic of métiers – and technical efficiency are likely to vary continuously, such that no two fishers are identical. However, there is empirical evidence that métiers can be grouped somewhat discretely into groups of vessels with highly similar relative catch rates (Branch et al. 2005), and it is possible that the same is true for efficiency. For example, vessels of similar size with similar distances between ports of origin and fishing grounds may have similar fuel and labour costs. Thus, the fleet diversity that is important in practice is likely to be the number of these different semi-discrete vessel 'types', and more importantly, how different they are from one another in their efficiency and relative catch rates. Fishers in different métiers may sometimes face different prices for the same catch. For example, fish caught in passive gears (e.g. traps, longlines) are often higher quality (e.g. FAO 2002), and thus sometimes fetch higher prices than fish caught in active gears (e.g. trawls, seines). However, differences in prices faced by different fishing units that are proportionally consistent across stocks (e.g. fishing unit $j$ receives prices twice as high as fishing unit $k$ for all stocks) are mathematically equivalent to differences in efficiency, and thus should not impact the results.

Recent studies of specific fisheries provide some support for the benefits of diversity in métiers in both managed and unmanaged contexts. For example, Dougherty et al. (2013) showed, theoretically and in simulations of western U.S. groundfish fisheries, that setting multispecies harvest quotas at a local scale to achieve coast-wide goals could increase overall fishery yields without increasing the likelihood of collapsing any of the stocks. If métiers vary spatially due to the varying juxtapositions of species' ranges on different fishing grounds, the type of management Dougherty et al. (2013) propose is equivalent to regulating the relative efforts in different métiers. Similar results have been found in several other multispecies fisheries (Sanchirico et al. 2006; Marchal et al. 2011;

Ulrich et al. 2011; Sibert et al. 2012). Burgess et al. (2013) provide evidence suggesting that the diversification of métiers in Western and Central Pacific tuna fisheries resulting from the expansion of industrial deep-set longline, purse-seine and pole-and-line fishing methods reduced the long-term threat of collapse posed by these fisheries to many tuna and billfish populations by tending to increase the balance in their exploitation rates.

There is also some empirical support for the ideas that fishers with high technical efficiencies can competitively dominate diverse fleets and increase overfishing, and that management can exploit differences among fishers' technical efficiencies to increase fishery-wide profits. For example, Schnier and Felthoven (2013) found that the introduction of individual transferable quotas (ITQ) in the Bering Sea and Aleutian Island crab fishery increased the likelihood of inefficient fishers exiting the fishery. This is consistent with theory suggesting that rights-based fishery management leads to fishery-wide efficiency by incentivizing the redistribution of fishing effort to more efficient fishers (Grafton et al. 2000). Similar results have been seen in other fisheries (e.g. Weninger 1998; 2008; Brandt 2007; Costello et al. 2010; Grainger and Costello 2012). In unmanaged fisheries, the rise of industrial fishing in the mid-$20^{th}$ century and subsequent fishery collapses (e.g. Myers and Worm 2003; Pauly et al. 2005; Worm et al. 2006; 2009) may be a consequence of the ability of efficient fishers to competitively dominate fisheries, and the increases in ecological risk associated with efficiency gains.

Though the results presented here suggest that management is likely to face fewer tradeoffs between yield- or profit-maximization and species conservation in diverse fleets, it is important to note that the structures of some food webs may make such tradeoffs difficult or impossible to overcome, regardless of fleet diversity. For example, if a fishery targets both a predator and its prey, maximizing yield or profit across the whole fishery might require the elimination of the predator to increase prey catches. Matsuda and Abrams (2006) explore such tradeoffs in detail and outline several instructive examples. Similarly, as discussed in Appendix D, diversification of either métiers or efficiency in unmanaged fisheries may exacerbate indirect threats from fisheries to some specialist predators and mutualists. The vulnerability of top predators and other marine species to fishing-induced trophic cascades has been documented both empirically (e.g. Estes et al. 1998; Jackson et al. 2001; Frank et al. 2005; Myers et al. 2007) and theoretically (e.g. May et al. 1979). Ecosystem-based fishery management (Pikitch et al. 2004) or other holistic approaches to fishery management may be particularly important in large fisheries with diverse fleets.

The focus of my analysis of unmanaged fisheries on equilibrium statics rather than dynamics did not allow the analysis to consider the possibility of transient stock collapses, effects of fleet diversity on the temporal variance in yields and profits, or the stability of fished ecosystems. Fleet diversity likely plays an important role in determining the stability of fishing yields and profits, and fished stocks. Ecological theory suggests that increasing biological diversity decreases the stability of individual species' populations (May 1973), but increases the stability of aggregate ecosystem services such as productivity (Lehman and Tilman 2000). This suggests that increasing fleet diversity may analogously destabilize the populations of individual stocks or the profits of fishers in individual fishing units, but may increase the stability of fishery-wide yields and profits. Recent evidence from California Current fisheries suggests that fisheries can indeed destabilize individual fish populations (Hsieh et al. 2006; Anderson et al. 2008). By decreasing the stability of individual stocks, increasing fleet diversity may also increase the likelihood of transient stock collapses. The effects of fleet diversity on the stability of the economic and ecological impacts of fisheries merit further study.

Fisheries are an important global provider of food, employment, and other social benefits (Beddington et al. 2007; Worm et al. 2009; Costello et al. 2012a; Chan et al. 2012), but also have large and increasing ecological impacts (Worm et al. 2006, 2009; Costello et al. 2012b; Halpern et al. 2012; Ricard et al. 2012). With global fish demand rising (Delgado et al. 2003) and global protein demand expected to double in the next half-century (Tilman et al. 2011), fisheries management faces the delicate challenge of providing the highest possible levels of sustainable production at the lowest possible ecological cost. The theory presented here, for which there is some empirical support in the literature, broadly suggests that diversifying métiers can have positive impacts on the yields and ecological sustainability of both managed and unmanaged multispecies fisheries. My analysis also suggests that the potential of management to improve fisheries' socio-economic and ecological outcomes relative to unmanaged outcomes is likely to be highest in diverse fishing fleets. Large international fishing fleets targeting widespread or migratory stocks, such as those targeting tunas, are likely to be diverse, but are also some of the most difficult to manage (Beddington et al. 2007; Worm et al. 2009). Continued advances in the management of these fisheries will be critical to ensuring the long-term sustainability of the socioeconomic benefits of fisheries and the ecosystems that support them. In the meantime, continuing to diversify métiers in multispecies fisheries may reduce threats to some weak stocks even without management (e.g. Burgess et al. 2013).

**Acknowledgements:** This work was supported by the Natural Sciences and Engineering Research Council of Canada (NSERC PGSD3-389196-2010) and by a University of Minnesota Doctoral Dissertation Fellowship. I thank Paul Venturelli, David Tilman, Stephen Polasky, Peter Abrams, Forest Isbell, Paul Marchal, Laura Dee, and several anonymous reviewers for helpful comments on earlier versions of the manuscript.

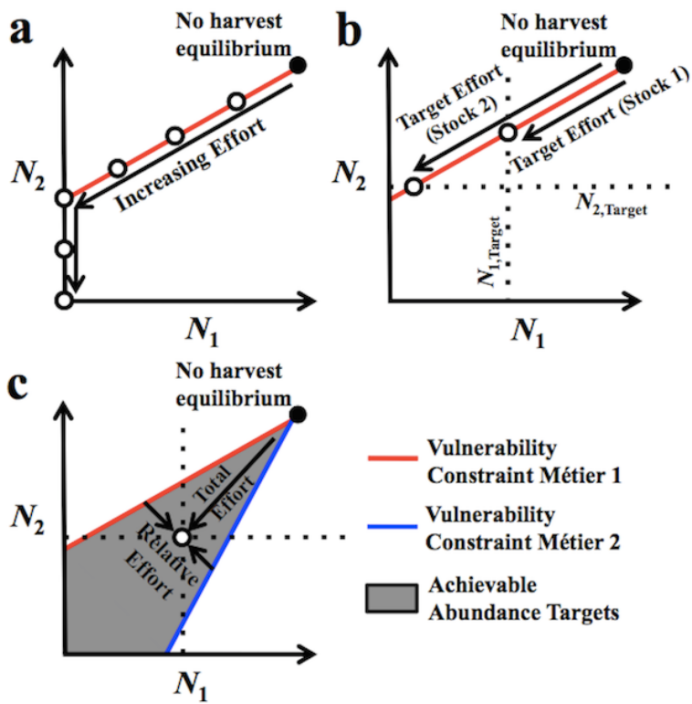

**Figure 1. Total and relative efforts and achievable abundances in managed fisheries.**
(**a**) With a single métier in a fishery, stocks' relative equilibrium abundances are constrained by their relative vulnerabilities to the métier. The 'vulnerability constraint' of a métier (red line for Métier 1) is the set of possible equilibrium abundances with no other métiers present in the fishery. Open circles represent some of these possible equilibrium abundances. (**b**) As a result, with a single métier (or few relative to the number of stocks) in a multispecies fishery it is often impossible to simultaneously achieve target abundances for multiple stocks, instead trading off overexploiting some with under-exploiting others, as illustrated by the open circles representing the equilibria resulting from harvesting one of the stocks at its target abundance. (**c**) With multiple métiers, management influencing both their relative and total effort levels can produce any set of equilibrium stock abundances within the region (shaded) bounded by the vulnerability constraints each would produce in isolation (red and blue lines). Increasing diversity of métiers increases the chances of this region including the target set of abundances (shown as an open circle).

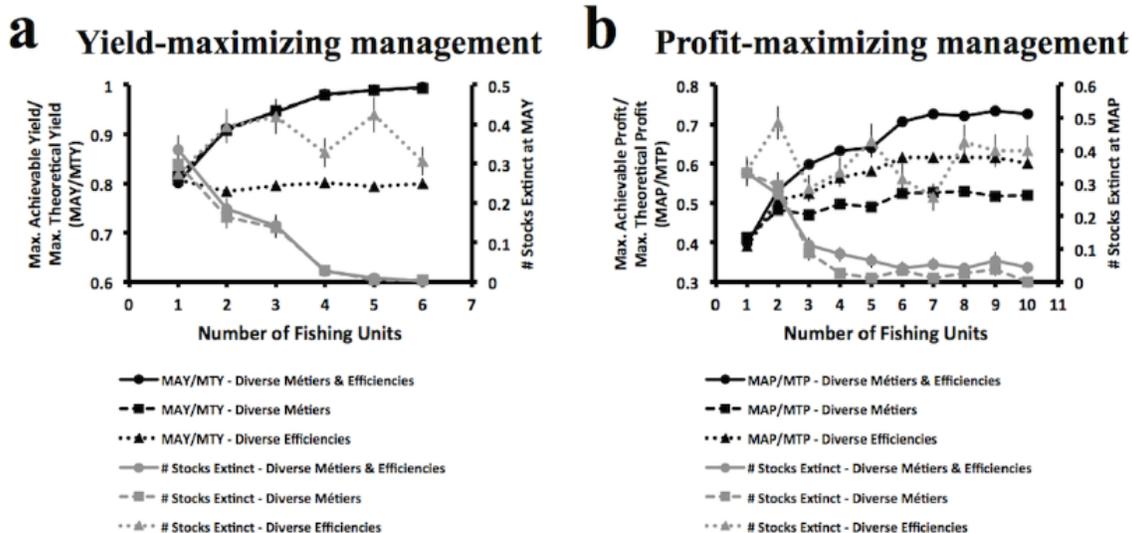

**Figure 2. Fleet diversity increases achievable yields and profits in managed fisheries.** Results of a 5-stock stochastic simulation of average relationships between fleet diversity and achievable yields and profits with (**a**) optimal management for obtaining maximum achievable yield (MAY) and (**b**) optimal management for obtaining maximum achievable profit (MAP). Average maximum achievable (**a**) yields (MAY) and (**b**) profits (MAP) are shown (black) along with the average number of extinctions achieving MAY or MAP requires (grey). Each point represents a sample of 500 models with randomly chosen parameter values. Vertical lines indicate standard errors. Note: points in (**a**) showing simulation results with diverse métiers only (black squares, black dashed line), and those showing results with diversity in both métiers and efficiency (black circles, black solid line) are nearly identical. This is due to the fact that diversity in efficiency among fishing units has no effect on achievable yields.

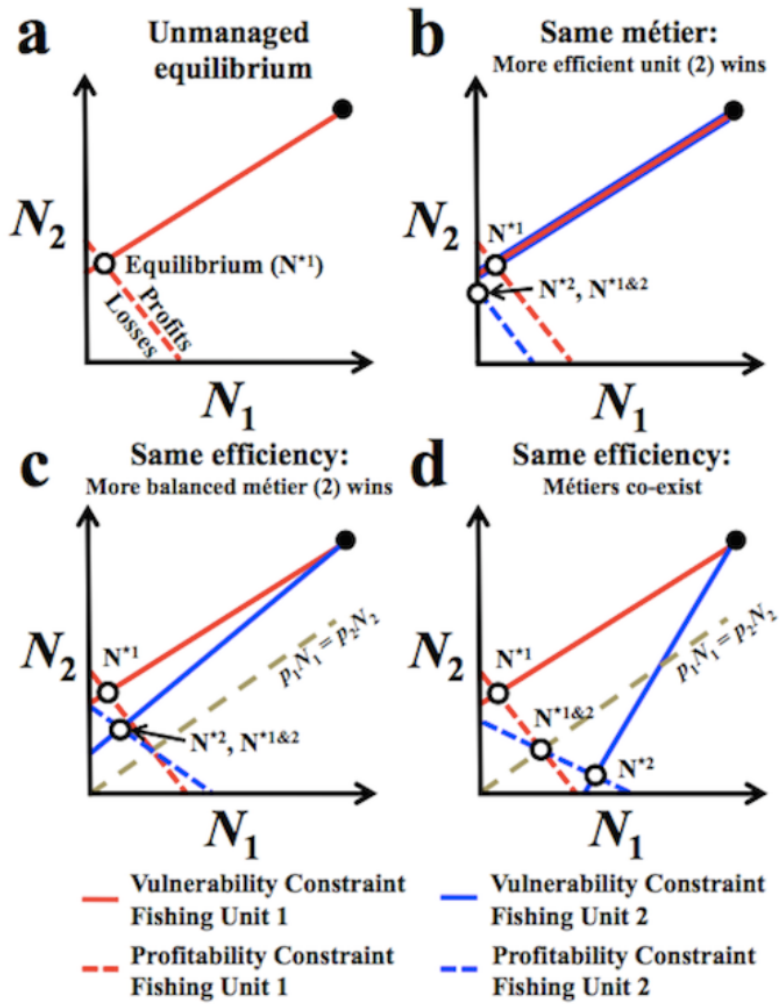

**Figure 3. Competition in unmanaged fisheries.** (**a**) With one fishing unit, equilibrium occurs at the intersection of the vulnerability constraint and profitability constraints ($N^{*1}$ for fishing unit 1). (**b**) If fishers are in the same métier, more efficient fishers will outcompete less efficient fishers (competitive equilibrium is denoted $N^{*1\&2}$). (**c**), (**d**) As métiers diversify with equal efficiency, equilibrium abundances will tend toward equalization of (price x abundance) is equal for all stocks (tan dashed line), either via (**c**) competitive exclusion, (**d**) co-existence, or both. The profitability constraints of all fishing units with equal efficiency (red and blue dashed lines in (**c**) and (**d**)) intersect at a single point where (price x abundance) is equal for all stocks (as drawn, $p_2 > p_1$).

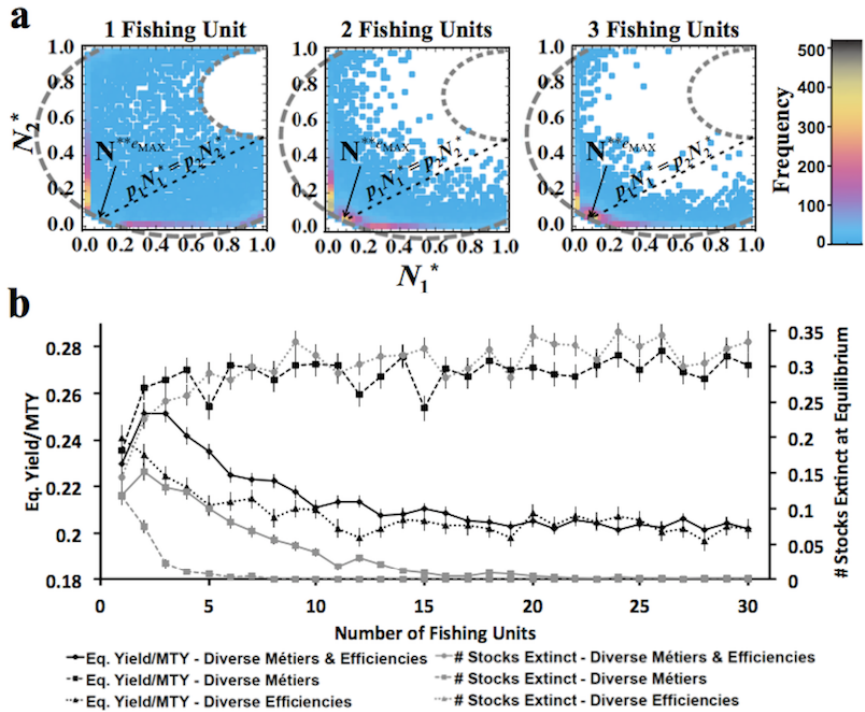

**Figure 4. Consequences of fleet diversification in unmanaged fisheries.** (**a**) A stochastic simulation showing the frequency distribution of equilibria ($\mathbf{N}^*$) in a 2 stock fishery in which 1, 2, and 3 fishing units respectively (from left to right) were randomly drawn 15 000 times from a uniform distribution of all possible métiers ($m_{ij} \sim U[0,1]$) and a bounded range of efficiencies ($e_j \sim U[1,12]$). Stocks were assumed to have logistic growth with $r_i$, $K_i = 1$ for $i = 1, 2$ and $p_1 = 1$, $p_2 = 2$. Grey lines represent all possible intersections of vulnerability and profitability constraints for the most (large dashes) and least (small dashes) efficient fishers. As a result of competition among fishers, increasing fleet diversity increases causes the probability of having the most efficient possible exploitation and stock abundances where (price x abundance) are equal at equilibrium ($\mathbf{N}^{**e_{MAX}}$, intersection of grey and black dashed lines) approaches 1. (**b**) A stochastic simulation with the same parameter values and distributions as (**a**), where each of 1 to 30 fishing units are drawn randomly 1000 times, with métiers and efficiencies varying separately and jointly. When only métiers vary, diversification leads to higher yields (black) and fewer extinctions (grey) on average by balancing exploitation. When only efficiencies vary, diversification has the opposite effect, by reducing equilibrium abundances. When both métiers and efficiencies vary, the probability density of equilibrium abundances concentrates at $\mathbf{N}^{**e_{MAX}}$ as diversity increases. As a result, the effect of métiers' diversity reducing extinctions dominates (grey), but in this case the effect of efficiencies reducing average equilibrium yields also dominates (black). Error bars shown are standard errors ($n = 1000$).

**Appendices for:**

**Consequences of fleet diversification in managed and unmanaged fisheries**

Matthew G. Burgess

**Appendix A. Vulnerability and profitability constraints under different ecological and economic assumptions**

*Vulnerability constraints*

The vulnerability constraint of a particular fishing unit, $j$, is the set of possible equilibrium abundance vectors, $\mathbf{N}^*$, at different efforts with only unit $j$ in the fishery. The vulnerability constraint is mathematically defined by the system of equations (3) for each stock. From this system of equations, it follows that any pair of extant stocks, $x$ and $y$, must satisfy the following at equilibrium:

$$\frac{g_x(\mathbf{N}^*)}{g_y(\mathbf{N}^*)} = \frac{m_{xj}}{m_{yj}} \quad (A.1)$$

As can be seen from equation (A.1), the vulnerability constraint is determined solely by the métier, and is not influenced by efficiency. Additionally, because it is defined by a system of $S - 1$ unique equations, it will be a one dimensional curve with any number of stocks, $S$. Figure A1a-e illustrates vulnerability constraints in 2-stock models under different ecological assumptions (i.e. values of $g_i(.)$), listed below.

*Figure A1a – No interaction*: $g_i(\mathbf{N}(t)) = r_i(1 - (N_i(t)/K_i))$, for $i = 1, 2$, where $r_i$ and $K_i$ are positive constants respectively representing the maximum intrinsic per-capita growth rate and carrying capacity of stock $i$ (sensu Schaefer 1954).

*Figure A1b – Competition*: $g_i(\mathbf{N}(t)) = r_i\left(1 - \dfrac{(N_i(t) + \alpha_{ij}N_j(t))}{K_i}\right)$, for $i, j = 1, 2, j \neq i$, where

$r_i$ and $K_i$ are the same as above, and $\alpha_{ij}$ is a positive constant representing the strength of the competitive effect of individuals of stock $j$ on the growth rate of stock $i$ (sensu MacArthur and Levins 1967). Stable co-existence of both stocks in the absence of harvesting requires $\alpha_{ij} \leq (K_i/K_j)$ for $i, j = 1, 2$.

*Figure A1c – Mutualism*: $g_i(\mathbf{N}(t)) = r_i\left(1 - \dfrac{(N_i(t) - \alpha_{ij}N_j(t))}{K_i}\right)$, for $i, j = 1, 2, j \neq i$, where $r_i$

and $K_i$ are the same as above, and $\alpha_{ij}$ is a positive constant representing the strength of the mutualistic effect of individuals of stock $j$ on the growth rate of stock $i$.

*Figure A1d – Predator and non-essential prey*:

$$g_{\text{pred}}(\mathbf{N}(t)) = r_{\text{pred}}\left(1 - \dfrac{(N_{\text{pred}}(t) - \alpha_{\text{predprey}}N_{\text{prey}}(t))}{K_{\text{pred}}}\right),$$

$$g_{\text{prey}}(\mathbf{N}(t)) = r_{\text{prey}}\left(1 - \dfrac{(N_{\text{prey}}(t) + \alpha_{\text{preypred}}N_{\text{pred}}(t))}{K_{\text{prey}}}\right),$$ where $r_i$ and $K_i$ are the same as above ($i$

= pred, prey), and $\alpha_{\text{predprey}}$ and $\alpha_{\text{preypred}}$ are positive constants respectively representing the strength of the positive effect of prey on predator growth rates and predators on prey growth rates. Note that the predator does not require any particular abundance of prey to survive (with no prey, the predator has logistic growth).

*Figure A1e – Predator and essential prey*: $g_{\text{pred}}(\mathbf{N}(t)) = r_{\text{pred}}\left[\dfrac{(N_{\text{prey}}(t) - N_{\text{prey,MINpred}})}{(K_{\text{prey}} - N_{\text{prey,MINpred}})}\right]$,

$g_{\text{prey}}(\mathbf{N}(t)) = r_{\text{prey}}\left(1 - \dfrac{(N_{\text{prey}}(t) + \alpha_{\text{preypred}}N_{\text{pred}}(t))}{K_{\text{prey}}}\right)$, where the prey growth is the same as in Figure A1d, but the predator now needs a minimum abundance of prey, $N_{prey,MINpred}$ to survive.

As illustrated in Figure A1 and in the main text for the logistic model (equation (4)), the shape of the vulnerability constraint is largely determined by the type of interaction (predation, competition, etc.), and by relative values of ($m_{ij}/r_i$) in most communities.

*Profitability constraints*

The profitability constraint of a particular fishing unit, $j$, is the set of abundance vectors, **N**, that result in zero profits for fishers in unit $j$. It is defined mathematically by setting the right hand side of equation 2b equal to 0 (equivalent to equation (5)):

$$e_j \sum_{i=1}^{S} p_i(N_i) m_{ij} N_i = 1 \quad (A.2).$$

As can be seen from equation (A.2), the profitability constraint is determined by both the métier (its slope; Figure 3c,d) and the efficiency (its position relative to the origin; Figure 3b) of fishers in fishing unit $j$. It is an $S - 1$ dimensional surface, which is linear if prices are constant (Figure 3). If prices increase as abundance decreases ($p_i'(.) < 0$), but not fast enough to cause profits to increase with decreasing abundance (i.e. $d[p_i(N_i)N_i]/dN_i > 0$ for all $N_i$, $i$), the profitability constraint is generally convex, as illustrated for a 2-stock fishery in Figure A1f, because the decrease in price as the abundance of a stock increases generally results in diminishing marginal returns to stock abundance (i.e. $d^2[p_i(N_i)N_i]/dN_i^2 < 0$ for all $N_i$, $i$). This is analogous to the property of convex isoquants as a result of diminishing marginal returns to production factors (e.g. capital, labour) in microeconomics (see Mas-Colell et al. 1995).

If a stock, *i*, has the property that the revenues it generates increase as its abundance decreases (i.e. $d[p_i(N_i)N_i]/dN_i < 0$), then it is likely to be driven extinct (Courchamp et al. 2006), and this likelihood can be exacerbated by increasing the diversity of métiers. If $d[p_i(N_i)N_i]/dN_i < 0$ at current and all lower abundances, $N_i$, of stock *i*, then any fishing unit whose current revenues from only stock *i* are greater than its costs can profitably harvest stock *i* to extinction. Increasing the diversity of either efficiency or métiers would increase the likelihood of including such a fishing unit by chance. If $d[p_i(N_i)N_i]/dN_i < 0$ at current abundance of stock *i* but not abundances lower than a certain abundance (because of price saturation, for example), increasing the diversity of efficiency or métiers would increase the likelihood of including a fishing unit that could drive stock *i* to this abundance, at which point the theory presented in the main text would apply, and equilibrium at infinite fleet diversity would satisfy equation (6). Thus, equation (6) (equality of (price x abundance) across all extant revenue-generating stocks at equilibrium) is likely to be satisfied even if prices can increase faster than abundances decline for some stocks.

**Appendix B. Fleet diversity in managed fisheries: Expanded mathematical treatment**

Consider the model of an *S*-stock, *J*-fishing unit fishery from the main text, where equation (2c) describes the population growth of stock *i*:

$$\frac{dN_i(t)}{dt} = N_i(t)\left(g_i(\mathbf{N}(t)) - \sum_{j=1}^{J} m_{ij} e_j E_j(t)\right) \quad (2c),$$

and equation (2b) describes the profits of fishing unit *j* at time *t*:

$$\pi_j(t) = E_j(t)\left(e_j \sum_{i=1}^{S} p_i(N_i(t)) m_{ij} N_i(t) - 1\right) \quad (2b).$$

In addition, it is instructive to formally partition fishing effort into total and relative efforts, by representing effort in fishing unit $j$, $E_j(t)$ at time $t$ (which is in monetary units), as the total cost-budget of the fishery as a whole (total effort), denoted $E(t)$ at time $t$, multiplied by the fraction of this cost-budget allocated to fishing unit $j$ (relative effort), denoted $b_j(t)$ at time $t$, where $\sum_j b_j(t) = 1$. Thus, $E_j(t) = E(t)b_j(t)$. With this substitution, the fishery-wide yield (the sum of catch-rates across stocks and fishing units) at time $t$, $Y(t)$ is given by:

$$Y(t) = \sum_{i=1}^{S} N_i(t) F_i(t) \quad (B.1a),$$

where $F_i(t)$, the fishing mortality rate of stock $i$ at time $t$, is given by:

$$F_i(t) = E(t) \sum_{j=1}^{J} b_j(t) e_j m_{ij} \quad (B.1b).$$

The fishery-wide profits at time $t$, $\Pi(t)$, are given by:

$$\Pi(t) = \left[\sum_{i=1}^{S} N_i(t) F_i(t) p_i(N_i(t))\right] - E(t) \quad (B.1c).$$

At any equilibrium, $\{\mathbf{N}^*, \mathbf{E}^*\}$ ($E^* = \{E_1^*, \ldots, E_J^*\}$), the population size of stock $i$ cannot be changing (i.e. $\left[(dN_i(t)/dt)|_{N_i(t)=N_i^*}\right] = 0$ for all $i$) (by definition of equilibrium), which implies that harvest is equal to surplus production,

$$N_i^* g_i(\mathbf{N}^*) = N_i^* \sum_{j=1}^{J} m_{ij} e_j E_j^* \quad (B.2).$$

This implies that equilibrium fishery-wide yields and profits can be expressed in terms of surplus production as:

$$Y^* = \sum_{i=1}^{S} N_i^* g_i(\mathbf{N}^*) \quad (B.3a).$$

$$\Pi^* = \sum_{i=1}^{S} p_i(N_i^*) N_i^* g_i(\mathbf{N}^*) - E^* \quad \text{(B.3b)}.$$

As can be seen from equation (B.3a), equilibrium fishery wide yield ($Y^*$) depends on the equilibrium stock sizes ($\mathbf{N}^*$), but not the total fishing effort (i.e. cost), $E^*$, needed to achieve these. In contrast, equilibrium fishery-wide profit depends on both the equilibrium stock sizes ($\mathbf{N}^*$), which determine yield and prices, and the total effort ($E^*$), which is equal to the total cost because effort is defined in units of monetary cost throughout this analysis. Moreover, every set of equilibrium stock abundances, $\mathbf{N}^*$, corresponds to a unique set of fishing mortality rates, $\mathbf{F}^* = \{F_1^*,\ldots, F_S^*\}$. The uniqueness of $\mathbf{F}^*$ given $\mathbf{N}^*$ follows from equation (B.2).

The set of fishing mortality rates at time $t$, $\mathbf{F}(t)$, is given by:

$$\mathbf{F}(t) = E(t)(\mathbf{M} \bullet \boldsymbol{\beta}(t)) \quad \text{(B.4)},$$

where $\mathbf{M}$ is a matrix defining the set of métiers in the fishery, $\mathbf{M} = \{\{m_{11},\ldots,m_{1J}\},\ldots,\{m_{S1},\ldots,m_{SJ}\}\}$, and $\boldsymbol{\beta}(t)$ is a vector of the relative efforts in different fishing units ($b$) weighted by their efficiencies $\boldsymbol{\beta}(t) = \{b_1(t)e_1,\ldots,b_J(t)e_J\}$. It is possible for both $E(t)$ and $\boldsymbol{\beta}(t)$ to be fully controlled by management under any set of efficiencies, $\{e_j\}_{\forall j}$, provided all efficiencies are positive and there are no limits on total effort, $E(t)$. Thus, the space of possible fishing mortality combinations, $\mathbf{F}$, and by extension, equilibrium abundance combinations, $\mathbf{N}^*$, is defined solely by the set of métiers in the fishery (defined by the matrix $\mathbf{M}$).

Consequently, the maximum achievable yield (MAY) in a fishery, which is determined only by the space of achievable abundance combinations (from equation B.3a), is influenced by the available diversity of métiers in the fishery, but not by the available diversity of efficiencies. Increasing the number of different métiers increases the space of possible equilibrium abundance combinations, which increases the likelihood that the highest yielding (or revenue generating) abundance combinations will be achievable. In fact, the addition of a new fishing unit, $k$, with a unique métier to a fishery cannot reduce the maximum yield (MAY) or revenue achievable through management, because any equilibrium abundance combination, $\mathbf{N}^*$, which was achievable before the introduction of fishing unit $k$ would still be achievable by setting $b_k = 0$. By similar logic, increasing the diversity of métiers cannot possibly worsen tradeoffs between overexploiting some stocks and underexploiting others (because additional diversity in $\mathbf{M}$ cannot shrink the space of achievable abundances), and increasing diversity in efficiency has no effect on the existence of such tradeoffs. Thus, increasing the diversity of métiers in a managed fishery should lead to increases in achievable yields and revenues (and profits by extension) and lessen the ecological costs of yield or profit-maximization. In contrast, increasing the diversity of efficiencies should have no effect on achievable yields or the tradeoffs between yield- or profit-maximization and conservation. The simulation results in Figure 2 support these predictions.

The achievability of a particular target set of abundances, denoted $\mathbf{N}_T$ (e.g. $N_{MSY}$ for all stocks, resulting in the maximum theoretical yield (MTY) fishery-wide), is unlikely unless there are at least as many métiers as stocks. Achieving the set of fishing mortality rates, $\mathbf{F}_T$, associated with $\mathbf{N}_T$ requires total ($E_T$) and relative ($\mathbf{b}_T$, $\boldsymbol{\beta}_T$) efforts solving:

$$\mathbf{F}_T = E_T \left( \mathbf{M} \bullet \boldsymbol{\beta}_T \right) \text{ (B.5)}.$$

An exact solution to this equation requires $\mathbf{M}$ to have a rank of $S$ or greater, which requires at least $S$ different métiers. Figure B1 illustrates this graphically in a 3-stock fishery. When the dimensionality of the set of achievable abundance targets (determined by the number of different métiers) is smaller than the number of stocks, it is highly unlikely that a particular set of target abundances will be contained in this set (Figure B1a,b vs. c).

Though not impacting achievable yields or yield-conservation tradeoffs, the available diversity of efficiencies does impact the achievable fishery-wide profits. Diversity in efficiency creates opportunities to increase profits by redistributing fishing effort from low-efficiency fishing units to high efficiency fishing units. Moreover, increasing diversity in efficiencies also increases the likelihood of including high efficiency extremes, which causes fisheries with higher diversity in efficiency to have higher maximum achievable profit (MAP) on average, holding the diversity of métiers constant.

As an illustrative example, suppose all fishing units have the same métier (i.e. $m_{ij} = m_{ik}$ for all $i,j,k$), denoted **m** (**m** = $\{m_1,\ldots, m_S\}$), and the efficiency of each fishing unit is drawn from a uniform distribution, $e_j \sim U[e_{MIN}, e_{MAX}]$. In this fishery, the maximum achievable profit (MAP) results from allocating all of the fishing effort to the most efficient fishing unit. The expected value of the efficiency of the most efficient of $J$ fishing units with efficiencies drawn randomly from $U[e_{MIN}, e_{MAX}]$ is given by:

$$E\left[\max\{e_j\}_{j=1}^{J}\right] = e_{MIN} + (e_{MAX} - e_{MIN})\left(\frac{J}{J+1}\right) \quad (B.6),$$

which is increasing in the number of fishing units ($J$), and approaches $e_{MAX}$ as $J$ approaches infinity. Moreover, given an existing set of $J$ fishing units, all from métier **m** with efficiencies $\{e_j\}$, making an additional fishing unit, $k$, available to the fishery cannot decrease the maximum available efficiency (i.e. the largest $e_j$ among all fishing units), and the probability that it will increase the maximum available efficiency will be positive (equal to $\Pr[e_k] > \max\{e_j\}$, which will be $1 - (J/(1+J))$ on average if the efficiencies of original $J$ units and the additional unit, $k$, are drawn from the same uniform distribution). Thus, the maximum achievable profit (MAP), which is determined by $\max\{e_j\}$, increases on average as the diversity of efficiencies increases. The simulation results in Figure 2 also support this prediction.

**Appendix C. Fleet diversity in unmanaged fisheries: Expanded proof of equations (6), (7), and (8), and note on the impossibility of priority effects**

The analysis of fleet diversity in unmanaged fisheries in this study is again based on the general model of $S$ stocks and $J$ fishing units initially in the fishery, in which the dynamics of each stock, $i$, are described by equation (2c) and the profits of each fishing unit are described by equation (2b):

$$\frac{dN_i(t)}{dt} = N_i(t)\left(g_i(\mathbf{N}(t)) - \sum_{j=1}^{J} m_{ij} e_j E_j(t)\right) \quad (2c),$$

$$\pi_j(t) = E_j(t)\left(e_j \sum_{i=1}^{S} p_i(N_i(t)) m_{ij} N_i(t) - 1\right) \quad (2b).$$

As stated in the main text, I also assume that the rate of change of fishing effort in fishing unit $j$, $dE_j(t)/dt$, is positive if $\pi_j(t)$ is positive, $dE_j(t)/dt < 0$ if $\pi_j(t) < 0$, and $dE_j(t)/dt = 0$ if $\pi_j(t) = 0$. At equilibrium (by definition), all $dE_j(t)/dt = 0$ (which implies that all $\pi_j(t) = 0$) and all $dN_i(t)/dt = 0$. This means that equilibrium with a single fishing unit, $j$, in the fishery occurs at a set of abundances, denoted $\mathbf{N}^{*j}$ in the main text, at the intersection of the vulnerability constraint (the set of abundances where $dN_i(t)/dt$ can be equal to 0 for all $i$ when only fishing unit $j$ is present in the fishery) and the profitability constraint (the set of abundances where $\pi_j(t) = 0$) of fishing unit $j$ – a point which was denoted $\mathbf{N}^{*j}$ in the main text. It also implies that equilibrium with multiple fishing units occurs at the intersection of all profitability constraints of fishing units with positive equilibrium effort ($E_j^* > 0$), because $\pi_j^* = 0$ for all fishing units, $j$, present.

*Competition favours efficiency*

If all fishing units have the same métier (i.e. $m_{ij} = m_{ik}$ for all $i$, $j$, $k$), then the fishing unit with the highest efficiency ($e_j$ for fishing unit $j$) will always have the highest profits, by equation (2b). Because $dE_j(t)/dt > 0$ if $\pi_j(t) > 0$, $dE_j(t)/dt < 0$ if $\pi_j(t) < 0$, and $dE_j(t)/dt = 0$ if $\pi_j(t) = 0$, as is assumed, the fishing unit with the highest efficiency will outcompete all others and be the only one at equilibrium. It will increase in effort until stocks are depleted to abundances at which it has zero profits, which would imply negative profits, and thus declining effort for all other fishing units. By the same logic, it is impossible for a fishing unit to persist in a fishery in which another fishing unit exists with the same métier and a higher efficiency.

As illustrated in Appendix B above, if the efficiencies of fishing units in a particular fishery are drawn randomly from a distribution with a fixed upper bound, denoted $e_{MAX}$, then the expected efficiency of the most efficient fishing unit in the fishery increases as the number of fishing units in the fishery increases (e.g. according to equation (S.B.6) if the underlying distribution is $U[e_{MIN}, e_{MAX}]$), and the probability of the fishery including a fishing unit with efficiency $e_{MAX}$ approaches 1 as the number of fishing units in the fishery approaches infinity. Similarly, if there is a set of possible métiers from which métiers of individual fishing units are drawn, the probability of the fishery including a fishing unit with efficiency $e_{MAX}$ and any particular métier also approaches 1 as the number of fishing units approaches infinity. Because, no fishing unit can persist in a fishery with another with greater efficiency and from the same métier, the probability that all stably persisting fishing units have an efficiency of exactly $e_{MAX}$ approaches 1 as the number of fishing units initially in the fishery approaches infinity.

*Competition promotes balanced exploitation*

In the main text, it was asserted that, if all métiers are technologically feasible (i.e. the métier of each fishing unit in the fishery is drawn randomly from a distribution with positive probabilities for all $0 \leq m_{ij} \leq 1$), and efficiencies are drawn from a distribution in which the maximum efficiency, $e_{MAX}$, is the same for all métiers, then any stock, $i$, extant at an equilibrium resulting from infinite initial fleet diversity has an abundance, $N_i^{**e_{MAX}}$, given by equation (7):

$$p_i\left(N_i^{**e_{MAX}}\right) N_i^{**e_{MAX}} = {1}/{e_{MAX}} \quad (7).$$

This result follows from two other results. The first of these is the above-demonstrated result that, when starting from an infinitely diverse fleet, all extant fishing units at equilibrium must have the maximum possible efficiency, $e_{MAX}$. The second result is that of equation (6), which states that, in a fishery with an initially infinite number of randomly selected fishing units, all having the same efficiency, the following relationship (equation (6)) must hold for the equilibrium abundances of any two stocks, $x$ and $y$, extant at any resulting equilibrium:

$$p_x\left(N_x^*\right)N_x^* = p_y\left(N_y^*\right)N_y^* \quad (6).$$

At any point, **N**, satisfying equation (6), all possible fishing units with the same efficiency (e.g. $e_{MAX}$) have the same per-unit-effort profits (from equation (2b)),

$$\pi_j(\mathbf{N}) = E_j(t)\left[e_{MAX} p_i(N_i) N_i - 1\right] \quad (C.1),$$

where $i$ and $j$ could respectively be any extant stock (by equation (6)) and fishing unit. Thus (by equations (5) and (C.1)), the profitability constraints of all possible fishing units with efficiency $e_{MAX}$ intersect at the point $\mathbf{N}^{**e_{MAX}}$ described by equation (7); and more generally, all possible fishing units having an identical efficiency (regardless of what the shared efficiency is) have profitability constraints intersecting at a point satisfying equation (6).

In addition, it is impossible to reach an equilibrium that does not satisfy equation (5), starting from an infinite number of fishing units with the same efficiency (e.g. $e_{MAX}$). As demonstrated in the main text, for any equilibrium with any number of fishing units, all having identical efficiency, $e_{MAX}$, that does not satisfy equation (6), there exists at least one métier that would have positive profits at that equilibrium, and thus be able to invade and disrupt the equilibrium. The likelihood of this métier being included in the fishery would approach 1 as the initial number of fishing units approached infinity.

To summarize, with infinite initial fleet diversity in a fishery, only fishing units with the maximum possible efficiency ($e_{MAX}$) can persist, and the only possible equilibrium point is $\mathbf{N}^{**e_{MAX}}$ from equation (7). Increasing diversity in efficiency pushes the aggregate efficiency of the fishery towards $e_{MAX}$, and increasing diversity in métiers pushes the fishery towards equality in (price x abundance) among extant stocks, assuming all métiers are feasible, and $e_{MAX}$ is the same for all métiers. Equation (8) is derived identically to equation (7), but under a relaxation of this latter assumption, whereby there is now an efficiency weight, ($a_i$ for stock $i$) associated with each stock, designed to capture stock-specific differences in catchability.

*Note: Priority effects on the outcome of competition require different fishing units to face oppositely differing prices for the same catch of at least one pair of stocks*

Priority effects occur when two fishing units compete and neither can invade the other's equilibrium, resulting in an outcome of competition determined by which unit enters the fishery earlier or can expand faster. Priority effects are impossible if all fishers face the same prices. More specifically, for a priority effect to occur between a pair of fishing units, $j$ and $k$, there must be at least one pair of stocks, $x$ and $y$, for which prices received by fishers in units $j$ and $k$ differ oppositely (i.e. $p_{xj} > p_{xk}$ and $p_{yj} < p_{yk}$, or vice versa, where $p_{ab}$ denotes the price received for an individual caught of stock $a$ by fishers in fishing unit $b$). This property is illustrated in Figure C1 in a model of a fishery targeting two stocks with logistic growth.

Analogously to priority effects between two ecological consumers (see Tilman 1980), priority effects between two fishing units result in the existence of an unstable equilibrium, where the profitability constraints of both fishing units are satisfied, and additional effort in each fishing unit would shift stock abundances away from the equilibrium to levels where it made higher profits than the other fishing unit (Figure C1). This implies that for priority effects to exist between two competing fishing units, $j$ and $k$, there must be at least one stock, $x$, for which $e_j m_{xj} > e_k m_{xk}$ and $p_{xj} e_j m_{xj} < p_{xk} e_k m_{xk}$ (i.e. more effort in fishing unit $j$ would reduce the abundance of stock $x$ relative to others, which would have a greater negative impact on fishing unit $k$'s revenues than its own), and similarly there must be at least one stock, $y$, for which $e_j m_{yj} < e_k m_{yk}$ and $p_{yj} e_j m_{yj} > p_{yk} e_k m_{yk}$. This implies that $p_{xj} < p_{xk}$ and $p_{yj} > p_{yk}$, and therefore $x$ and $y$ must be separate stocks.

**Appendix D. Exceptions to common effects of fleet diversification on yield and ecological impacts of unmanaged fisheries, as a result of ecology or technological feasibility**

Provided efficiency is finite, infinitely diverse métiers in a multispecies fishery drive stocks' abundances to a point in the first quadrant where they generate equal marginal revenue (equation (8)), preventing extinction of weak stocks directly caused by the fishery. This property is illustrated in Figure 4 in a model with no interspecific interactions in Figure 4, but also holds under many types of interactions. For example, Figure D1a shows the results of a similar stochastic simulation of a fishery targeting two competing stocks, with the same qualitative results as in Figure 4b. The procedure is the same as in Figure 4b, except the population growth of both stocks in the absence of fishing is described by a simplified Lotka-Volterra competition model (sensu MacArthur and Levins 1967), where $g_i(\mathbf{N}(t)) = 1 - N_i(t) - 0.3 N_j(t)$ ($i, j = 1, 2, i \neq j$), and other parameter values/distributions are: $\{p_1 = 1, p_2 = 2, m_{1j} \sim U[0, 1], e_j \sim U[2, 12]\}$. As the number of fishing units increased, average yields decreased when efficiency varied, and increased when only métiers varied. The average number of extinctions decreased as the number of fishing units increased when métiers varied, and increased when only efficiency varied (Figure D1a, right panel).

However, some ecosystem structures or restrictions on the range of technologically feasible fishing units can cause diversification of métiers to lead to more frequent stock collapses. Ecological specialist stocks that either have obligate prey or mutualists also caught or otherwise impacted by the fishery may still be driven extinct. Specifically, a stock, $i$, having long-term persistence that requires an obligate mutualist or prey, $k$, to have at least a minimum population size, $N_{k,\text{MIN}i}$, will be driven extinct at infinite fleet diversity if $N_k^{**e_{\text{MAX}}} < N_{k,\text{MIN}i}$. This is illustrated in a stochastic simulation of a fishery targeting a predator (Stock 2) and its essential prey (Stock 1) in Figure D1b. The procedure was the same as in Figures 4b and D1a, except that stocks' population growth in the absence of fishing was now described by: $g_1(\mathbf{N}(t)) = 1 - N_1(t) - N_2(t)$,

$g_2(\mathbf{N}(t)) = \dfrac{(N_1(t) - N_{1,\text{MIN}2})}{(1 - N_{1,\text{MIN}2})}$, and other parameter values/distributions are: $\{p_1 = 1, p_2 = 0.5, m_{1j} \sim U[0, 1], e_j \sim U[2, 5]\}$. In this model, increases in all types of diversification led to increases in average likelihood of predator extinction (Figure D1b). This occurred because competition among diverse métiers and efficiency drives the prey's abundance to a level that is below $N_{1,\text{MIN}2}$ (Figure D1b, left panel). Additionally, all types of fleet diversification increased average yields, as the prey's average yields increased in response to reduced predation pressure, which more than compensated for lost predator yields (Figure D1b, right panel). This latter result is somewhat dependent on parameter values, but is likely to hold in systems where transfers of biomass up food chains are inefficient, a common property in nature (e.g. Lindeman 1942; Odum 1957; Christensen and Pauly 1992). Thus, diversification of métiers in fisheries impacting multiple trophic levels may increase both yields and the likelihood of stock collapses.

Diversification of métiers can also increase the likelihood of stock collapses when some relative catch rates are not technologically feasible. Some relative catch rates may not be feasible if, for example, two stocks have sufficiently high niche overlap that it would be difficult or impossible to design a fishing technology that catches one without also catching the other at a certain rate. If relative catch rates that lead stocks to have relative depletions satisfying equation (6) or (8) are not technologically feasible, then it is possible for métiers that drive one or more stocks extinct to be favoured by competition. Figure D1c illustrates this point in a stochastic simulation identical to Figure D1a, in which métiers for which $m_{1j} < 0.55$ are now technologically infeasible (i.e. $m_{1j} \sim$ U[0.55, 1]). As a result, the relationships between diversification in yield seen in Figure D1a are similar, but now all types of diversification increase the likelihood of stock 1's collapse (Figure D1c). An analytical example of this is also given below.

Suppose 2 stocks, $x$ and $y$, having logistic growth ($g_i(\mathbf{N}(t)) = r_i(1 - (N_i(t)/K_i))$ for all $i$) where $r_x = 2r_y$, $K_x = K_y = a_x = a_y = 1$, and $p_x = p_y = p$, are exploited in a fishery, and, due to technological constraints, $m_{xj} \leq m_{yj}$ for any fishing unit $j$. At any efficiency, $p_x N_x^*$ = $p_y N_y^*$ would require either a single fishing unit, $j$, where $m_{xj} = 2m_{yj}$, or 2 fishing units, $j$ and $k$, where $m_{xj} > 2m_{yj}$ and $m_{xk} < 2m_{yk}$, or vice versa. However, this is infeasible because $m_{xj} \leq m_{yj}$ for all $j$. Thus, $p_x N_x^* > p_y N_y^*$ at all feasible equilibria, implying that competition favours fishing units with the largest possible harvest rate of stock $x$, which in this case corresponds to $m_x = m_y$ (i.e. $m_x = m_y = 0.5$ because $m_{xj} + m_{yj} = 1$ for all $j$ by definition). Thus, if the maximum efficiency is $e_{MAX}$, infinite fleet diversity would result in equilibrium stock sizes, $N_x^* = (2/3pe_{MAX}) + (1/3)$, $N_y^* = (4/3pe_{MAX}) - (1/3)$. Infinite fleet diversity results in the extinction of stock $y$ if $e_{MAX} \geq 4/p$.

**References**

Christensen, V., and Pauly, D. 1992. ECOPATH II – a software for balancing steady-state ecosystem models and calculating network characteristics. Ecological Modelling **61**: 169-185.

Courchamp, F., Angulo, E., Rivalan, P., Hall, R.J., Signoret, L., Bull, L., and Meinard, Y. 2006. Rarity value and species extinction: The anthropogenic Allee effect. PLoS Biology **4**(12): e415. doi: 10.1371/journal.pbio.0040415.

Lindeman, R.L. 1942. The trophic-dynamic aspect of ecology. Ecology **23**: 399-418.

MacArthur, R., and Levins, R. 1967. The limiting similarity, convergence, and divergence of coexisting species. American Naturalist **101**: 377-385.

Mas-Colell, A., Whinston, M.D., and Green, J.R. 1995. Microeconomic theory. Oxford University Press, New York, NY.

Odum, H.T. 1957. Trophic structure and productivity of Silver, Springs, Florida. Ecological Monographs **27**(1): 55-112.
61

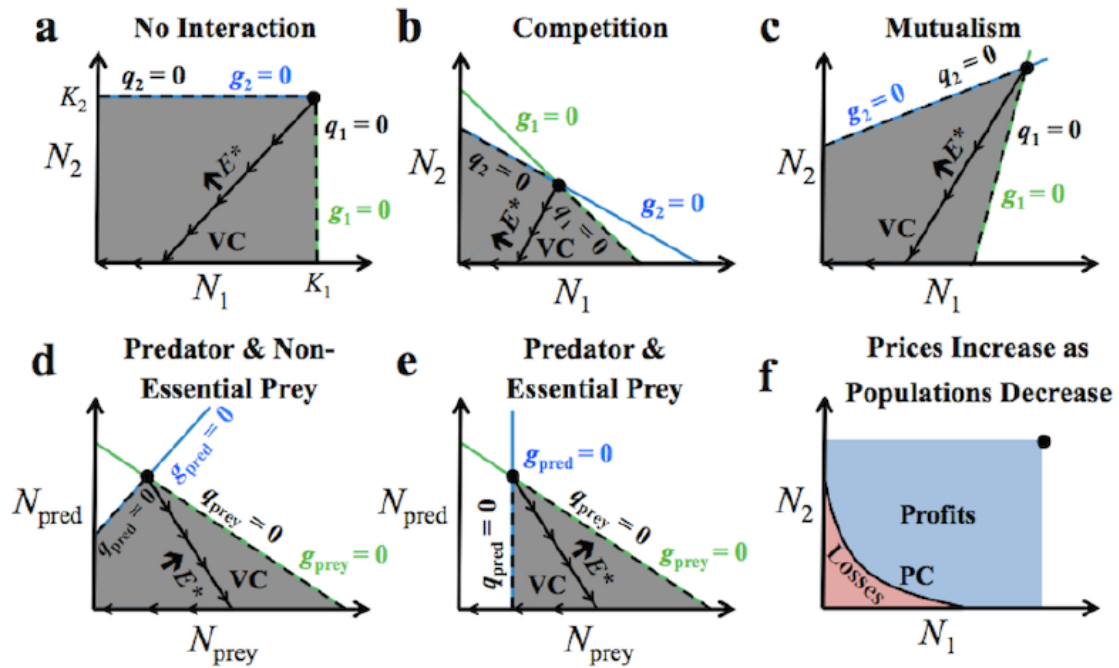

**Figure A1. Vulnerability and profitability constraints with different assumptions.** Vulnerability (**a-e**) (VC) and profitability constraint (**f**) (PC) with different ecological and economic assumptions. In each figure, stocks' zero net growth isoclines (ZNGI) ($g_i = 0$) (blue and green lines) and equilibrium abundances (filled circle) in the absence of harvesting are shown. If one of the stocks is not caught in the fishery ($q_i = 0$), increasing fishing effort causes equilibrium abundances to move along its ZNGI ($g_i = 0$) towards the origin (dashed lines). If both stocks are caught, increasing fishing effort causes equilibrium abundances to move along the vulnerability constraint, which must lie somewhere in the gray shaded region, and whose slope is determined by the stocks' relative catch rates and growth rates. Specific functional forms on which the shapes of ZNGIs are based for different classes of species interactions are given in Appendix A. (**f**) When prices increase as stocks' abundances decrease, the profitability constraint – the set of stock abundances that result in zero profits, separating abundances yielding positive profits (blue shaded region) and losses (red shaded region) – is generally convex.

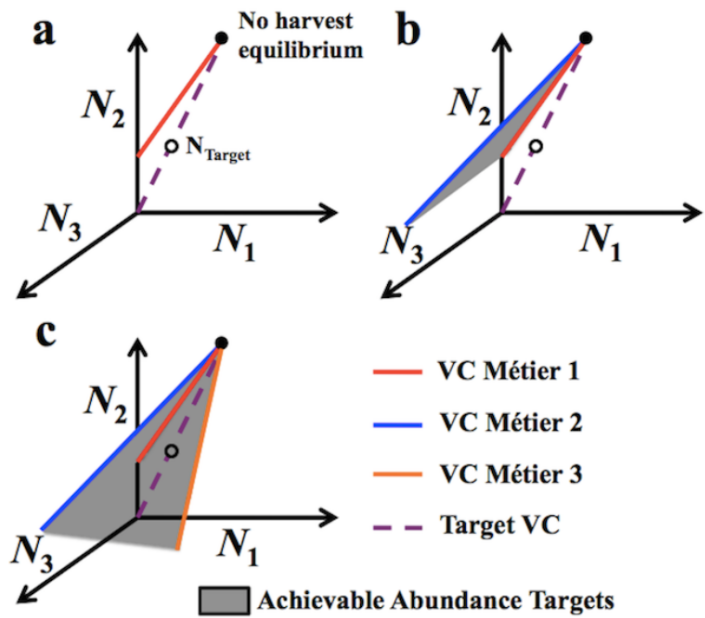

**Figure B1. Diversity of métiers and achievable abundances in managed fisheries.** The relationship between the number of métiers and the achievability of a multi-stock abundance target ($N_{Target}$, open circles) is shown in a 3-stock model. A hypothetical vulnerability constraint producing the target is also shown (dashed purple line). With only 1 métier (**a**), achievable outcomes are constrained to a single one-dimensional vulnerability constraint curve (red), which is unlikely to coincide exactly with the desired curve (purple) in 3-dimensional space by chance. Similarly, with 2 métiers (**b**) it is also unlikely, though less unlikely, that the 2-dimensional plane of possible achievable outcomes (shaded region) contains the desired outcome. However, with 3 (**c**) or more métiers, the chance that the desired outcome is achievable becomes sizeable, provided the 3 métiers differ in their relative catch rates.

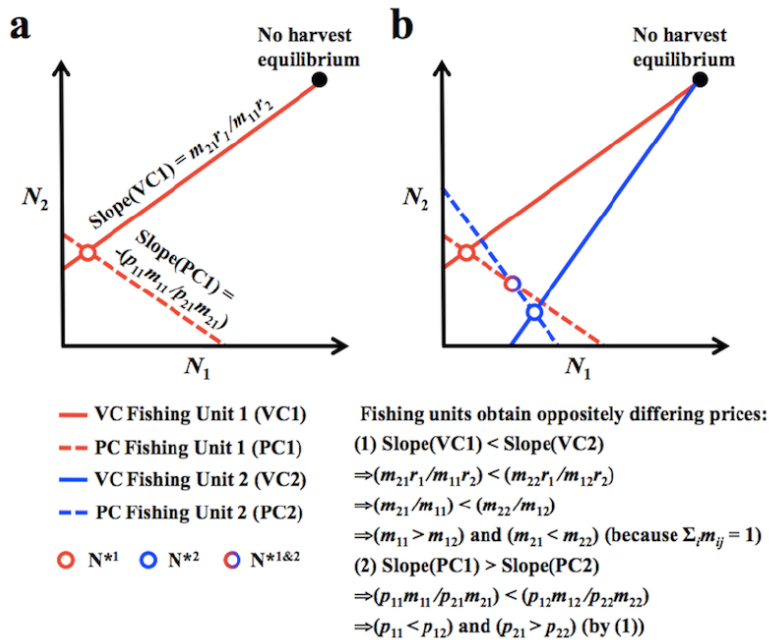

**Figure C1. Priority effects require fishing units to face oppositely differing prices for pairs of stocks.** A 2-stock 2-fishing unit model illustrating the reason for which priority effects are only possible if fishers in different fishing units face different prices. Slope equations for the vulnerability constraint (Slope(VC)) and profitability constraint (Slope(PC)) are derived respectively from equations (4) and (5), assuming there are 2 stocks, each having logistic growth and constant prices, where abundances ($N_i$, $i = 1,2$) are normalized as fractions of carrying capacity (i.e. $K_1 = K_2 = 1$). Circles indicate equilibrium stock sizes with: only fishing unit 1 ($\mathbf{N}^{*1}$, red), only fishing unit ($\mathbf{N}^{*2}$, blue), and the unstable co-existence equilibrium ($\mathbf{N}^{*1\&2}$, red and blue).

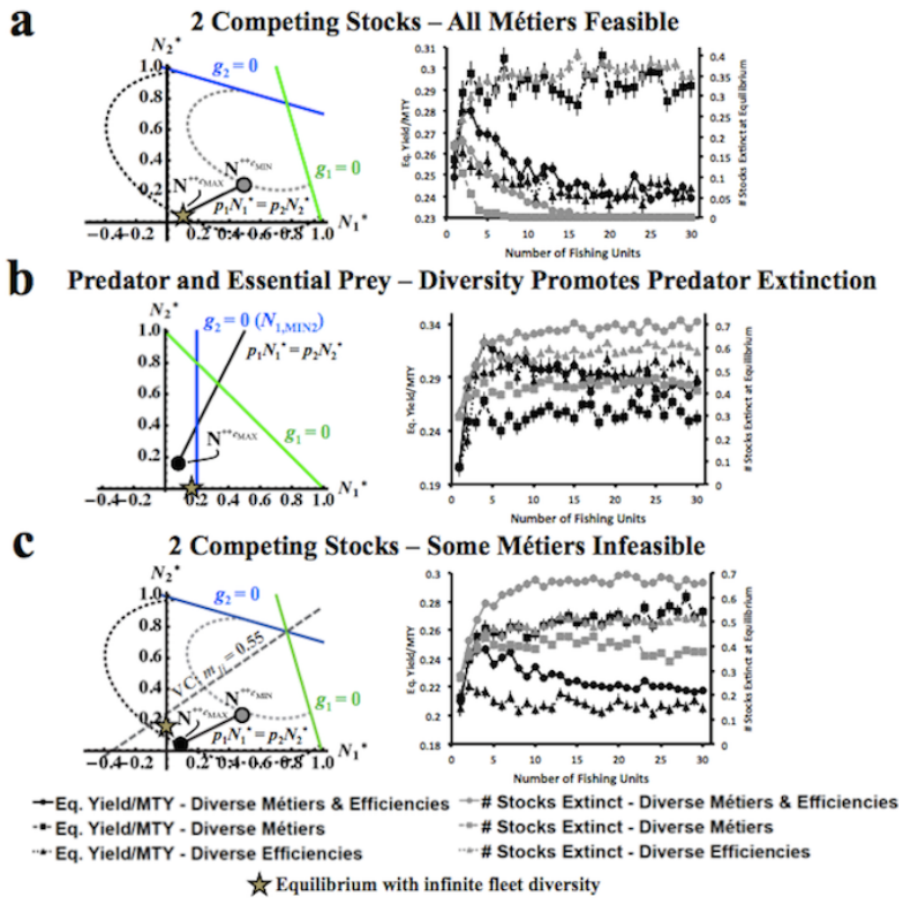

**Figure D1. Relationships between fleet diversity and yields and ecological impacts in different types of unmanaged fisheries.** These are illustrated in 2-stock fisheries in which (**a**, **c**) stocks are competing or are (**b**) predator (Stock 2) and essential prey (Stock 1). In panel (**c**), only métiers $j$ with $m_{1j} \geq 0.55$ are technologically feasible, illustrated by the grey dashed line in (**c**) (left). Points at which $p_1 N_1^* = p_2 N_2^*$ are illustrated (black solid lines), as well as all possible intersection points of vulnerability and profitability constraints for fishing units with minimum ($e_{MIN}$) (grey dotted lines) and maximum ($e_{MAX}$) (black dotted lines) feasible technological efficiency are shown in the left-hand panels. Stocks' zero net growth isoclines (ZNGI) ($g_i = 0$) (blue and green lines) are also shown. The right-hand panels show the relationships between fleet diversity and yield (black) and the average number of extinctions (grey) in stochastic simulations of the fisheries illustrated in the corresponding left-hand panels. Each point represents a sample of 1000 models with randomly chosen parameter values. Vertical lines indicate standard errors.